\documentstyle[floats,prl,aps,twocolumn,psfig]{revtex}

\def\xv{\vec{x}}

\def\Sz{S^z}
\def\abs#1{\vert #1 \vert}

\draft
\begin{document}

\twocolumn[\hsize\textwidth\columnwidth\hsize\csname @twocolumnfalse\endcsname
\title{Lanczos study of the $S=1/2$ frustrated square-lattice antiferromagnet
in a magnetic field}
\author{A.\ Honecker}
\address{Institut f\"ur Theoretische Physik, TU Braunschweig,
           Mendelssohnstr.\ 3, D--38106 Braunschweig, Germany}
\date{June 9, 2000; revised August 25, 2000}
\maketitle
\begin{abstract}
We study the zero-temperature phase diagram of the frustrated square-lattice
$S=1/2$ antiferromagnet in an external magnetic field numerically
with the Lanczos algorithm.
For strong frustration, we find disordered phases at high (and low)
magnetic fields. Between these two disordered phases we find a plateau
in the magnetization curve at half of the saturation
magnetization which corresponds to a state with up-up-up-down (uuud)
spin order. This and other considerations \cite{ZHP}
suggest an unusual ordering scenario: There are an ordered phase
with a spin gap (the plateau) and disordered magnetically gapless
phases above and below.

The transition to saturation is studied in further detail
and problematic conclusions in earlier investigations of this region
are pointed out.

\end{abstract}
]

\section{Introduction}

Frustrated magnets are known to exhibit rich physics and diverse properties
in zero external field (compare this volume and \cite{frust}).
Exciting behavior in an external magnetic field was also
observed recently in several cases: For instance, the $S=7/2$
three-dimensional antiferromagnet Gd$_3$Ga$_5$O$_{12}$
is disordered (but gapless) in zero external field
and orders only in finite magnetic fields \cite{GGG}.
Another example is the recently discovered two-dimensional
$S=1/2$ spin-gap material SrCu$_2$(BO$_3$)$_2$ which exhibits several
plateaux in the magnetization curve \cite{Kageyama}. The origin of these
plateaux and the nature of the corresponding spin states are under
intense debate \cite{theory} and it is suggested that some of the plateaux
in SrCu$_2$(BO$_3$)$_2$ might be described by certain types of ordered
states.

A plateau in the magnetization curve of the triangular lattice
at one third of the saturation magnetization is theoretically well
known (see, e.g., \cite{trian}). However, 
general magnetization plateaus in two dimensions (including those
of SrCu$_2$(BO$_3$)$_2$) are still not well understood
despite recent progress on the subject \cite{general2d}.

\section{The model}

\label{secMod}

In \cite{ZHP} we have studied the aforementioned issues in
one of the simplest two-dimensional frustrated magnets, the frustrated
square lattice antiferromagnet whose Hamiltonian is given by 
\begin{equation}
H = J \sum_{\langle \vec{x}, \vec{y} \rangle}
 \vec{S}_{\vec{x}} \cdot \vec{S}_{\vec{y}}
+ J' \sum_{[ \vec{x}, \vec{y} ]}
 \vec{S}_{\vec{x}} \cdot \vec{S}_{\vec{y}}
 - h \sum_{\vec{x}} S^z_{\vec{x}} \,
\label{H}
\end{equation}
where the first sum is over nearest neighbor pairs
$\langle \vec{x}, \vec{y} \rangle$ and the second one over diagonal
neighbor pairs $[ \vec{x}, \vec{y} ]$.
Here we will present details of \cite{ZHP} and further results
obtained from a numerical diagonalization of the $S=1/2$ version
of the Hamiltonian (\ref{H}) with the Lanczos method.

At zero field, the frustrated square lattice $S=1/2$ Heisenberg
model has been studied extensively using also exact
diagonalization \cite{ed,SchZi,SZP}.
One reason for the popularity of this model is that it
exhibits melting of long-range magnetic order by quantum fluctuations:
For small frustration $J' \ll J$, one finds antiferromagnetic
N\'eel order while for $J' \gg J$ a certain type of collinear
order described by a wave vector $(\pi,0)$ or $(0,\pi)$. Finally,
in an intermediate region around $J' = J/2$, a spin-liquid groundstate 
with a gap to magnetic excitations is found.

In the presence of an external field, an important quantity is the magnetization
\begin{equation}
\langle M \rangle = {1 \over S V} \left\langle
                    \sum_{\xv} \Sz_{\xv}\right\rangle
\label{defM}
\end{equation}
which we normalize to saturation values $\abs{\langle M \rangle} = 1$.
Magnetization curves for the model (\ref{H}) have already been
computed by exact diagonalization \cite{LoNo,YaMue}. However,
the study of
\cite{LoNo} was restricted to a special lattice of $4 \times 4$ sites
which can also be interpreted in terms of several other
geometries. For instance, a strip of width 4
is equivalent to a frustrated four-leg
ladder and then plateaux are expected in the magnetization curve
at $\langle M \rangle = 0$, $1/4$, $1/2$ and $3/4$ (see, e.g., \cite{CHP}).
Precisely these values were also observed in \cite{LoNo},
but it remains to be clarified whether they are an artifact of
the special geometry. The other study \cite{YaMue} excluded
the region $J' > J/2$ which we find to be the most interesting one
and boundary conditions were used in \cite{YaMue} which
frustrate the order which wants to develop.

\section{Single magnon dispersion}

\label{sec1Mag}

It is instructive to look first at the dispersion of a single magnon above
the ferromagnetic state with $\langle M \rangle = 1$ (all spins aligned
along the field). When we formally set $h=0$, the one-magnon dispersion
for the $S=1/2$ model is given by
\begin{eqnarray}
{\cal E}_{{\rm 1s}}(k_x,k_y) &=& - 2 (J + J')
 + J \left( \cos k_x + \cos k_y \right) \nonumber \\
&& + 2 J' \cos k_x \cos k_y \, .
\label{oneMag}
\end{eqnarray}
For $J' < J/2$, this dispersion has a single minimum at
$k_x = k_y = \pi$ corresponding to N\'eel order. On the other
hand, for $J' > J/2$ two equivalent minima are found at 
$k_x = 0$, $k_y = \pi$ or $k_x = \pi$, $k_y = 0$ which signal
collinear order. The case $J' = J/2$ is special: Here lines
of minima are found for either $k_x = \pi$ or $k_y = \pi$.

\section{Magnetization curves}

\begin{figure}[hbt]
\centerline{\psfig{file=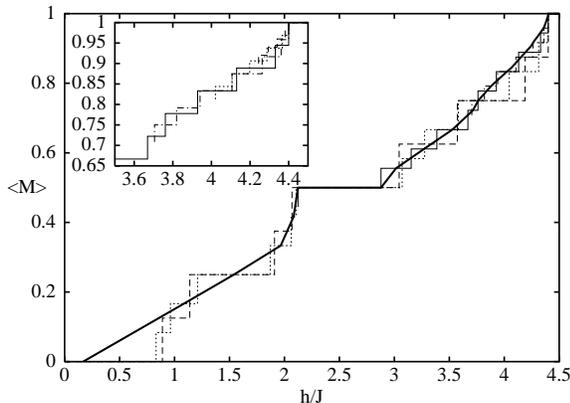,width=0.9\columnwidth,angle=270}}
\caption{
Zero-temperature magnetization curves of the $S=1/2$ model at $J'/J = 0.6$.
The lines are for clusters of size $4 \times 4$ (dashed), $6 \times 4$
(dotted), $6 \times 6$ (full) and $8\times 6$ (dashed-dotted)
(Inset: $8 \times 8$ (dotted) and $10 \times 10$ (dashed)).
The bold line is an extrapolation (compare the text for details).
}
\label{M0.6}
\end{figure}

We now discuss the behavior in the entire field range using
exact diagonalization. Guided by
the dispersion (\ref{oneMag}), we have considered only rectangular
clusters with an even number of sites in both directions.
Finite-size magnetization curves are shown in
Fig.\ \ref{M0.6} for $J'/J = 0.6$. As we have pointed out
in section \ref{secMod}, finite-size effects must be examined
carefully. Even if the $\langle M \rangle =  3/4$ and $1/4$ plateaux
are quite pronounced on the $4 \times 4$ and $6 \times 4$ clusters,
we believe that they are artifacts of the special system sizes.
Indeed, the data for wider strips indicates a vanishing plateau at
$\langle M \rangle = 3/4$ in the thermodynamic limit. In contrast,
the plateau at $\langle M \rangle = 1/2$ is only a
little narrower on the $6 \times 6$ cluster than for the smaller clusters.
In the extrapolation (bold line in Fig.\ \ref{M0.6}),
we have therefore drawn a plateau with $\langle M \rangle = 1/2$
but none at $\langle M\rangle = 1/4$ and $3/4$.
The value of the spin gap ({\it i.e.}\ the boundary
of the $\langle M\rangle = 0$ plateau) has been taken from \cite{KOSW}.
Otherwise the extrapolation was obtained by the standard
procedure of connecting the mid-points of the steps of the
magnetization curves at the largest available system size.

\begin{figure}[hbt]
\centerline{\psfig{file=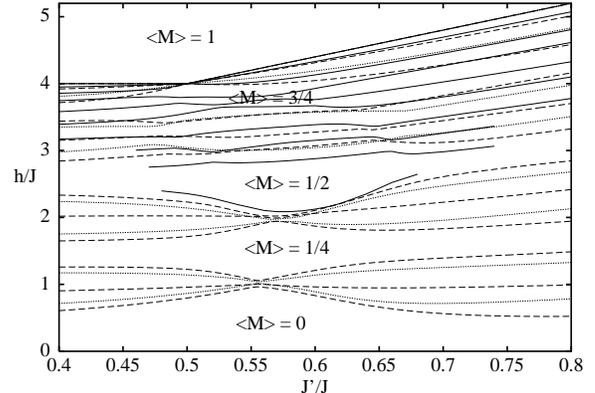,width=0.9\columnwidth,angle=270}}
\caption{
Transition lines for the $S=1/2$ model. The lines are for clusters
of size $4 \times 4$ (dotted), $6 \times 4$ (dashed) and
$6 \times 6$ (full).
}
\label{Trans}
\end{figure}

Fig.\ \ref{Trans} shows the projection of the magnetization curves
onto the horizontal axis, {\it i.e.}\ the locations of the finite-size
jumps in the magnetization curves. From this one can
then first read off how the width of the
$\langle M \rangle = 1/2$ plateau varies with $J'/J$.
An analysis of its finite-size behavior \cite{ZHP} indicates
that the $\langle M \rangle = 1/2$ plateau exists in the region
$0.5 \lesssim J'/J \lesssim 0.65$.

Fig.\ \ref{Trans} further shows that the transition to
saturation is special at $J' = J/2$, as is expected from
(\ref{oneMag}): A jump of size $\delta \langle M \rangle = 1/L$
occurs just below saturation where $L$ is the length of the shorter edge
of the cluster.
This implies a smooth transition to saturation in the thermodynamic
limit. The particularity of the transition to
saturation for $J' \approx J/2$ was already noted in \cite{YaMue}
where, however, the jump was obscured by the choice of cluster
geometries. We therefore believe the asymptotic form proposed
in \cite{YaMue} just to be a manifestation of these boundary effects
in combination with the questionable assumption that
$h(\langle M \rangle)$ has a power series expansion.
It should be possible to obtain the correct asymptotic
form from Bose condensation of magnons \cite{BC} into the lines of
minima of the single magnon dispersion (\ref{oneMag}), but this
computation has not yet been carried out.

For $J' \approx J/2$, the magnetization curves further show a
pronounced finite-size plateau at $\langle M \rangle = 1 - 1/L$,
{\it i.e.}\ just below the jump. It was actually proposed in
\cite{Gluzman} that a plateau just below saturation should
exist for $J' \gtrsim J/2$, but our candidate disappears as
$L \to \infty$ and we suspect an error in the treatment \cite{Gluzman} of
the single-magnon dispersion (\ref{oneMag}).

\section{Static structure factors}

\begin{figure}[hbt]
\centerline{\psfig{file=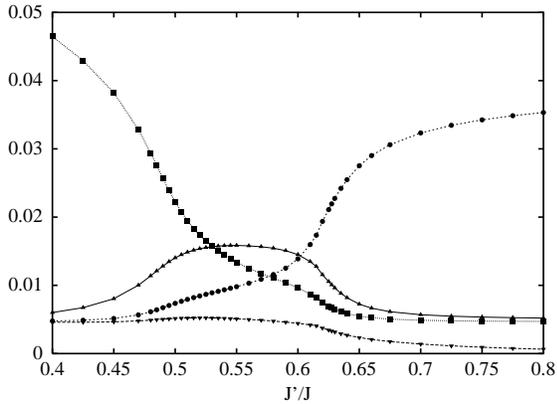,width=0.9\columnwidth,angle=270}}
\caption{
Peaks of the
static structure factors of the $S=1/2$ model as a function of
$J' / J$ at $\langle M\rangle = 2/3$ on the $6 \times 6$ cluster.
The longitudinal structure factor $S^{zz}(0,\pi)$ is shown
by triangles pointing up, $S^{zz}(\pi,\pi)$ by triangles pointing down
and the transverse structure factor $S^{xx}(0,\pi)$ by circles
and $S^{xx}(\pi,\pi)$ by squares.
\label{strucM2o3}
}
\end{figure}

\begin{figure}[hbt]
\centerline{\psfig{file=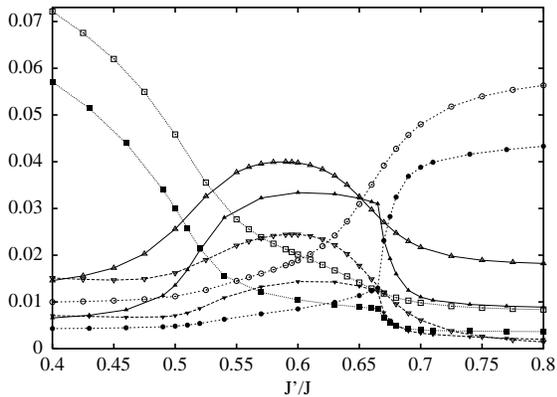,width=0.9\columnwidth,angle=270}}
\caption{
Same as Fig.\ \ref{strucM2o3} but for $\langle M\rangle = 1/2$ and
on the $6 \times 6$ cluster (filled symbols) as well as on the
$4 \times 4$ cluster (open symbols).
\label{strucM1o2}
}
\end{figure}

To gain further insight into the phase diagram and
in particular in order to understand the nature of the
$\langle M \rangle = 1/2$ plateau state better, we use
the static structure factors which are defined as\footnote{
The symmetrized variant
 $\left(S^{\alpha\beta}(k_x,k_y) + S^{\alpha\beta}(k_y,k_x)\right)/2$
is used
when the cluster is not square or the groundstate not symmetric under
reflection across the diagonal of the cluster ($x \leftrightarrow y$).}
\begin{equation}
S^{\alpha\beta}(k_x,k_y) = {1 \over V^2}
\sum_{x,y,r,s} {\rm e}^{i \left(k_x x + k_y y\right)}
\left\langle
S^{\alpha}_{r,s} S^{\beta}_{r+x,s+y} \right\rangle \, .
\label{DefStruc}
\end{equation}
The finite magnetization leads to a trivial peak in
$S^{zz}(0,0)$. We observe additional peaks in $S^{\alpha\beta}(k_x,k_y)$
at $(0,\pi)$ or $(\pi,0)$ and $(\pi,\pi)$. The evolution
of these peaks with $J'/J$ is shown in Figs.\ \ref{strucM2o3} and
\ref{strucM1o2} for two values of $\langle M\rangle$.

N\'eel and collinear order develops in the transverse components
for small and large $J'$, respectively. Just at $\langle M \rangle = 1/2$
and for $0.5 \lesssim J'/J \lesssim 0.65$, the longitudinal components develop 
uuud order with three spins pointing up and one down on a four-spin
plaquette. While it may be possible that some order exists in
$S^{zz}(0,\pi)$ at $\langle M \rangle = 2/3$
for intermediate $J'$ (see Fig.\ \ref{strucM2o3}), the structure
factors at other magnetizations and Monte Carlo simulations of the
classical model indicate disordered states for magnetizations sufficiently
far above or below the uuud state at $\langle M \rangle = 1/2$
\cite{ZHP}.

\section{Groundstate quantum numbers}

For an odd number of magnons $V/2-S^z$, one expects
the groundstate to carry $k_x = k_y = \pi$ (corresponding to
N\'eel order) for $J' \ll J$ while for $J' \gg J$ one expects
collinear order and thus $k_x = 0$, $k_y = \pi$ or $k_x = \pi$, $k_y = 0$
-- compare the discussion of the one-magnon sector in
section \ref{sec1Mag}.
When $V/2-S^z$ is even, the groundstate carries momentum
$k_x = k_y = 0$. However, as was pointed out in \cite{SchZi,SZP},
groundstate level crossings are still expected in the sectors with
$V/2-S^z = 2 n$ and $n$ odd. Then the groundstate is
even under diagonal reflections for $J' \ll J$ and odd for $J' \gg J$.

\begin{figure}[hbt]
\centerline{\psfig{file=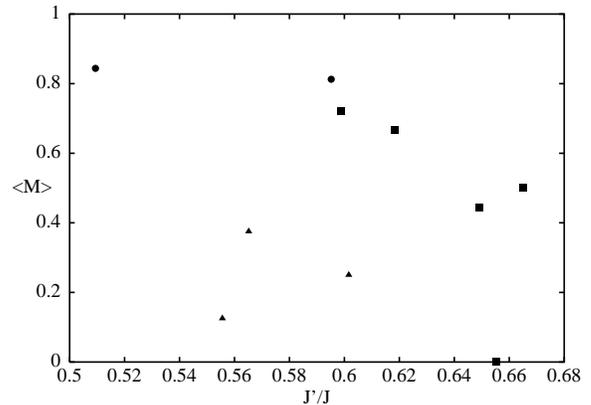,width=0.9\columnwidth,angle=270}}
\caption{
Positions of unique groundstate level crossings. Triangles are
for the $4 \times 4$ cluster, squares for the $6 \times 6$ cluster
and circles for the $8 \times 8$ cluster.
\label{gslcFig}
}
\end{figure}

The {\it unique} groundstate level crossings are shown in Fig.\ \ref{gslcFig}
(in some sectors more than one groundstate level crossing is observed).
The crossing at $\langle M \rangle = 0$ on the $6 \times 6$ cluster
was taken from \cite{SZP}.
Note that according to (\ref{oneMag}), the crossing point must move
to $J' = J/2$ for $\langle M \rangle \to 1$ where a first-order
transition between N\'eel order and collinear order occurs.
The position of the crossing at $\langle M \rangle = 1/2$ agrees
with our earlier estimates for the transition between the
uuud plateau state and collinear order. Also the crossing at
$\langle M \rangle = 0$ is close to recent estimates (see, e.g.,
\cite{KOSW}) for the location of the zero-field transition between
the disordered spin-liquid state and the collinear state.

\section{Summary}

We have studied the frustrated square lattice $S=1/2$ Heisenberg
model in the presence of
a strong magnetic field using the Lanczos method. A plateau with
$\langle M \rangle = 1/2$ exists in the region
$0.5 \lesssim J'/J \lesssim 0.65$ and corresponds to an
up-up-up-down ordered state.
The transverse spin components show N\'eel and collinear order
for small and large $J'$, respectively. The intermediate region
contains the ordered $\langle M \rangle = 1/2$ state and
disordered regions at higher and lower fields \cite{ZHP}.
Finally, we have found that the transition to saturation is special
at $J' = J/2$.

\bigskip
{\it Acknowledgments:}
I am indebted to O.A.\ Petrenko and M.E.\ Zhitomirsky for collaboration
on the subject \cite{ZHP}, to S.\ Gluzman and D.\ Poilblanc for discussions
and suggestions as well as to the Deutsche Forschungsgemeinschaft
for funding participation in the conference
`Highly Frustrated Magnetism 2000'.

\end{document}